\documentclass[]{spie}  
\usepackage[]{graphicx}
\usepackage[breaklinks, colorlinks, citecolor=blue, linkcolor=MyBlue, urlcolor=RoyalPurple, colorlinks=true, linkcolor=blue, debug, baseurl=' ']{hyperref}

\title{The NIKA 2012-2014 observing campaigns: control of systematic effects and results} 

\author{ 
A.~Catalano\supit{a},
R.~Adam\supit{a},
A.~Adane\supit{b},
P.~Ade\supit{c},
P.~Andr\'e\supit{d},
A.~Beelen\supit{e},
B.~Belier\supit{f},
A.~Beno\^it\supit{g},
A.~Bideaud\supit{c},
N.~Billot\supit{h},
N.~Boudou\supit{g},
O.~Bourrion\supit{a},
M.~Calvo\supit{g},
G.~Coiffard\supit{b},
B.~Comis\supit{a},
A.~D'Addabbo\supit{g,i},
F.-X.~D\'esert\supit{l},
S.~Doyle\supit{c},
J.~Goupy\supit{g},
C.~Kramer\supit{h},
S.~Leclercq\supit{b},
J.~F.~Mac\'ias-P\'erez\supit{a},
J.~Martino\supit{e},
P.~Mauskopf\supit{m,c},
F.~Mayet\supit{a},
A.~Monfardini\supit{g},
F.~Pajot\supit{e},
E.~Pascale\supit{c},
L. Perotto\supit{a},
E.~Pointecouteau\supit{o,p}
N.~Ponthieu\supit{l},
V.~Rev\'eret\supit{d},
A.~Ritacco\supit{a},
L.~Rodriguez\supit{d},
G.~Savini\supit{n},
K.~Schuster\supit{b},
A.~Sievers\supit{h},
C.~Tucker\supit{c} and
R.~Zylka\supit{b}
\skiplinehalf
\supit{a}Laboratoire de Physique Subatomique et de Cosmologie, Universit\'e Grenoble Alpes, CNRS/IN2P3, 53, rue des Martyrs, Grenoble, France \\
\supit{b}Institut de RadioAstronomie Millim\'etrique (IRAM), Grenoble, France \\
\supit{c}Astronomy Instrumentation Group, University of Cardiff, UK \\
\supit{d}Laboratoire AIM, CEA/IRFU, CNRS/INSU, Université Paris Diderot, CEA-Saclay, 91191 Gif-Sur-Yvette, France \\
\supit{e}Institut d'Astrophysique Spatiale (IAS), CNRS and Universit\'e Paris Sud, Orsay, France \\
\supit{f}Institut d'Electronique Fondamentale (IEF), Universit\'e Paris Sud, Orsay, France \\
\supit{g}Institut N\'eel, CNRS and Universit\'e de Grenoble, France \\
\supit{h}Institut de RadioAstronomie Millim\'etrique (IRAM), Granada, Spain \\
\supit{i}Universit\`a 'La Sapienza' di Roma, Italy \\
\supit{l}Institut de Plan\'etologie et dAstrophysique de Grenoble (IPAG), CNRS and Universit\'e de Grenoble, France \\
\supit{m}SESE and Physics, Arizona State University, Tempe, AZ, USA \\
\supit{n}University College London, Department of Physics and Astronomy, Gower Street, London WC1E 6BT, UK \\
\supit{o}Universit\'e de Toulouse, UPS-OMP, Institut de Recherche en Astrophysique et Plan\'etologie (IRAP), Toulouse, France \\
\supit{p}CNRS, IRAP, 9 Av. colonel Roche, BP 44346, F-31028 Toulouse cedex 4, France \\
}

\authorinfo{Further author information: (Send correspondence to Andrea Catalano)\\Andrea Catalano: E-mail: catalano@lpsc.in2p3.fr, Telephone: +33-476284152}

\begin{document} 
  \maketitle 

\begin{abstract}
The New IRAM KID Array (NIKA) is a dual-band camera operating with frequency multiplexed arrays of Lumped Element Kinetic Inductance Detectors (LEKIDs) cooled to 100 mK. NIKA is designed to observe the intensity and polarisation of the sky at 1.25 and 2.14~mm from the IRAM 30~m telescope. We present the improvements on the control of systematic effects and astrophysical results made during the last observation campaigns between 2012 and 2014. 
\end{abstract}


\section{INTRODUCTION}
\label{sec:intro}  

New challenges in millimetre wave astronomy require high sensitivity and high resolution instruments. As current detector technologies such as high impedance bolometers are currently photon noise limited for ground-based observations\cite{2012MmSAI..83...72T}, and have reached the limit of practical array sizes, gains in mapping speeds can only be achieved by moving to a new generation of detector suitable for use in large format arrays\cite{2010SPIE.7741E...2B, gismo, 2008ACT, 2008SPIE.7020E..25G}. One of the proposed technological solution is the Lumped Element Kinetic Inductance Detector (LEKID). LEKIDs allow for large multiplexing factors (up to 400 pixels)\cite{2012SPIE.8452E..0OB, Swenson2010} read out in the frequency domaine\cite{Bourrion2012}\ and simple manufacturing\cite{day2003}. This technological solution has been selected for the NIKA project that aims at constructing a dual-band millimeter camera for observations at 30~m IRAM telescope which is located in the Pico Veleta area of the Sierra Nevada, Spain, at an altitude of 2850~m\cite{NIKA2014,NIKA_2011}. This location is ideal to exploit the atmospheric windows centered at 2.1, 1.4, and 0.85~mm. The NIKA camera is few hundred-pixel demonstration instrument used as a pathfinder for future development of the final NIKA 2 camera and is the first successful demonstration of KID technology on a science grade telescope.The NIKA camera has been open to public observations since beginning 2014. The aims of the NIKA camera is to perform simultaneous observations in two millimeter bands (1.25~mm and 2.14~mm) of mJy point sources as well as to map extended continuum emission up to few arcmin in scale with diffraction limited resolution and background limited performance. Such observational requirements will lead the NIKA camera to be a science driver in several astrophysical fields as the follow-up of Planck satellite clusters via the Sunyaev-Zel'dovich effect, high redshift sources and quasars, early stages of star formation and nearby galaxies emission. In addition, the NIKA collaboration opened to host a polarised 1.25~mm array in the final NIKA2 instrument. The adopted solution uses a single birefringent sapphire plate to modulate the polarization of light incident on each array. This monochromatic Half Wave Plate (HWP) solution was preferred to a more sophisticated achromatic HWP to minimise and better control all the systematic errors linked to the plate taking into account the loss of polarisation performance. 

Previous observational campaigns had revealed several technical aspects that limited the sensitivity of the detectors and the overall calibration accuracy of the instrument. Some of the major aspects in particular were the real time optimisation of the detectors working point and the atmospheric absorption corrections that are crucial for reaching a high accuracy calibration. These improvements were implemented in the last observational campaigns and are presented in this paper. We also discuss how these technical improvements permitted us to reach astrophysical quality data by presenting results obtained for several astrophysical sources in intensity and in polarisation.
The current paper is structured as follows: in section \ref{sec:nika} we give the principal characteristics of the NIKA instrument. In section \ref{sec:cal} we present the NIKA calibration process  with particular interest on the improvements implemented with respect to the previous observational campaigns. In section \ref{sec:run5-6-7} we describe the principal results of the technical runs in November 2012, June 2013 and January 2014. In section \ref{sec:run8} we present the some preliminary results of the first open scientific pool performed in February 2014. Finally, in section \ref{sec:nika2} we describe the next generation NIKA2 instrument that will be installed for commissioning in 2015 at the 30~m IRAM telescope.

\section{The NIKA instrument state-of-the-art}\label{sec:nika}
The NIKA instrument comprises a cryostat with a closed-cycle $^3$He -$^4$He dilution refrigerator in order to reach a working temperature of about 100~mK, refractive cold optics to correctly coupled the instrument to the IRAM telescope, metal mesh filters to reject the unwanted emission of the sky and the telescope, two hundreds pixels KIDs arrays and the readout electronics\cite{NIKA2014}.
The principal characteristics of the NIKA instrument are listed in table \ref{table:car}. In the next subsections we detail the different sub-systems of the NIKA instrument.   

\begin{table}[h]
\caption{Principal characteristics of the NIKA instrument: The contribution of the atmosphere in the total background is derived considering good weather conditions. This corresponds to an expected photon noise level equal to about $5 \cdot 10^{-17}$ $W/\sqrt{Hz}$ at 2.14~mm channel and $9 \cdot 10^{-17}$ $W/\sqrt{Hz}$ at 1.25~mm channel.} 
\label{table:car}
\begin{center}       
\begin{tabular}{|l|l|l|} 
\hline
  & 1.25 mm & 2.14 mm \\
\hline
  Pixel Size [mm] & 1.6 &  2.3 \\
\hline
 Dual Polarisation & yes & yes \\
\hline
  Angular Size [$F_{lambda}$] & 0.9 & 0.79 \\
\hline
 Beam Efficiency [\%] &55 & 70 \\
\hline
 Detector Efficiency [\%] &80 & 80 \\
\hline
  Overall Optical Efficiency  [\%] & 30 & 30  \\
\hline
 Total Background  [pW] & 45 & 20 \\
\hline 
\end{tabular}
\end{center}
\end{table} 

\subsection{The cryostat}\label{ssec:cryostat}
Cooling the two KID arrays to 100~mK is the major requirement that drives the architecture of the NIKA instrument. This is achieved using a 4~K cryocooler and a closed-cycle $^3$He - $^4$He dilution. All the system can be easily operated and controlled from a remote station with very few a local interventions such as the switching on and
off of the pulse tube, and the refilling of the nitrogen carbon trap (used to prevent impurities into the mixture circuit). The final goal for NIKA2 is to be
able to fully remote-control the instrument. 

The NIKA detectors are sensitive to magnetic fields inside the cabin, due to the Earth and all the instrumentation present on site. In order to reduce these potential noise
sources, two magnetic shields have been added: mu-metal at 300~K and a superconducting lead screen on the 4K stage. 
In addition, the cryostat has been placed on a vibration reduction plate to suppress micro-phonics.


\begin{figure}
\begin{center}
\includegraphics[scale=0.2]{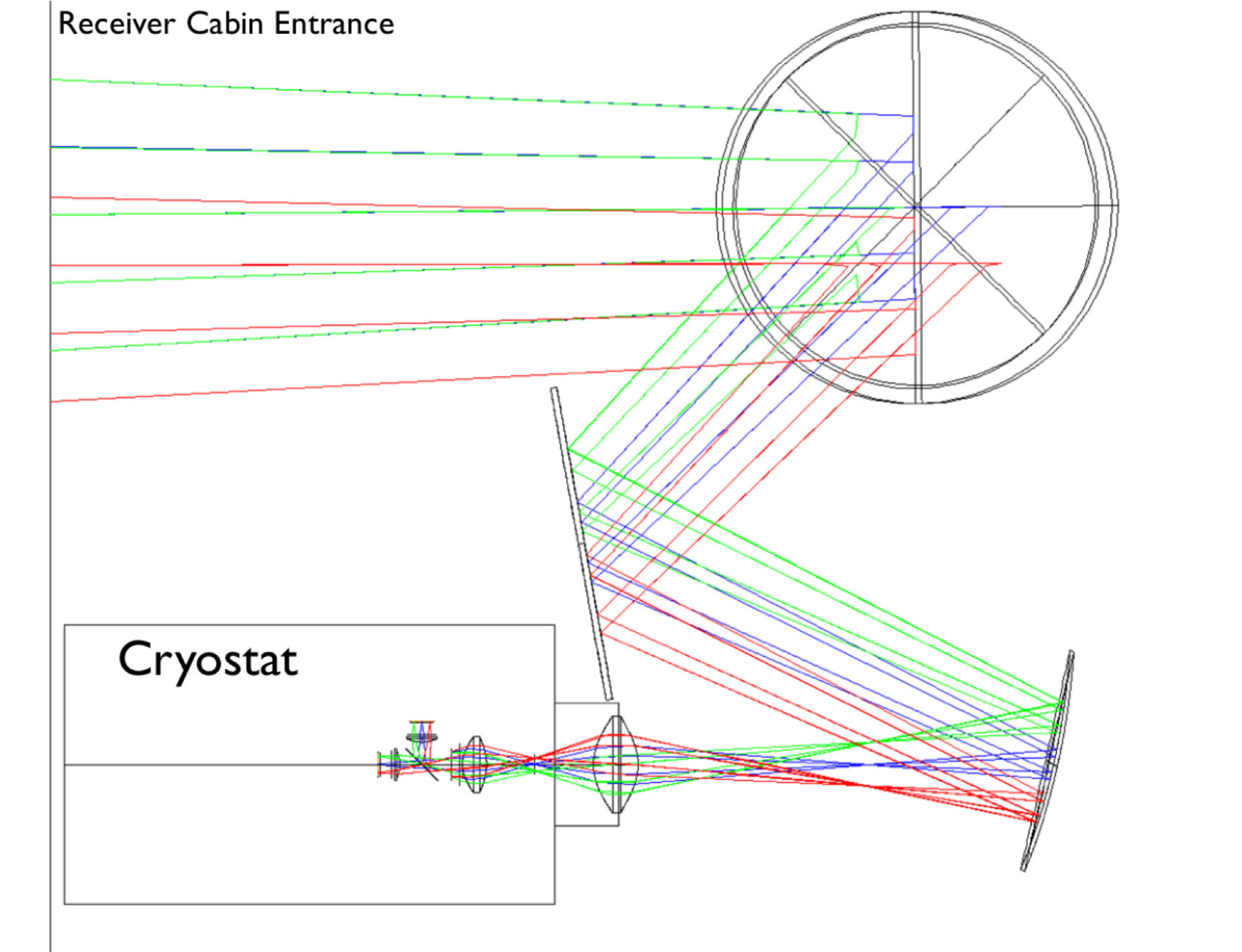}
\includegraphics[scale=0.25]{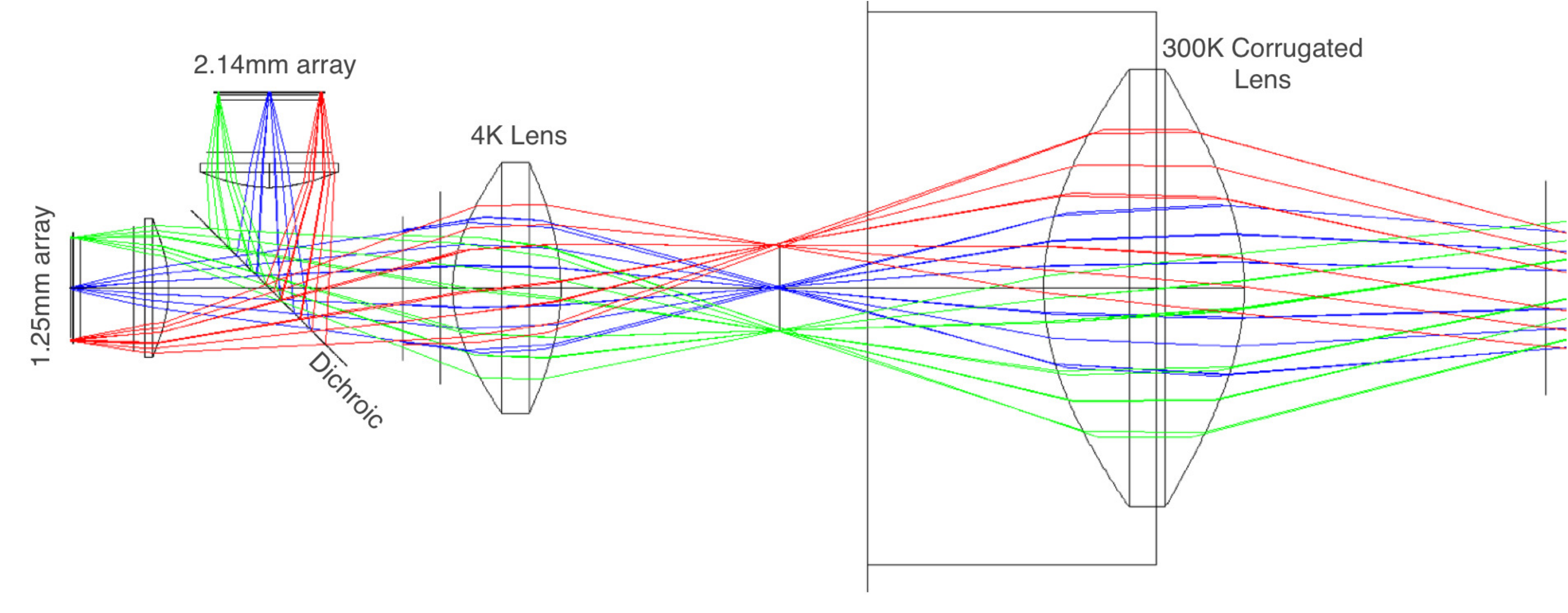} \\
\includegraphics[scale=0.2]{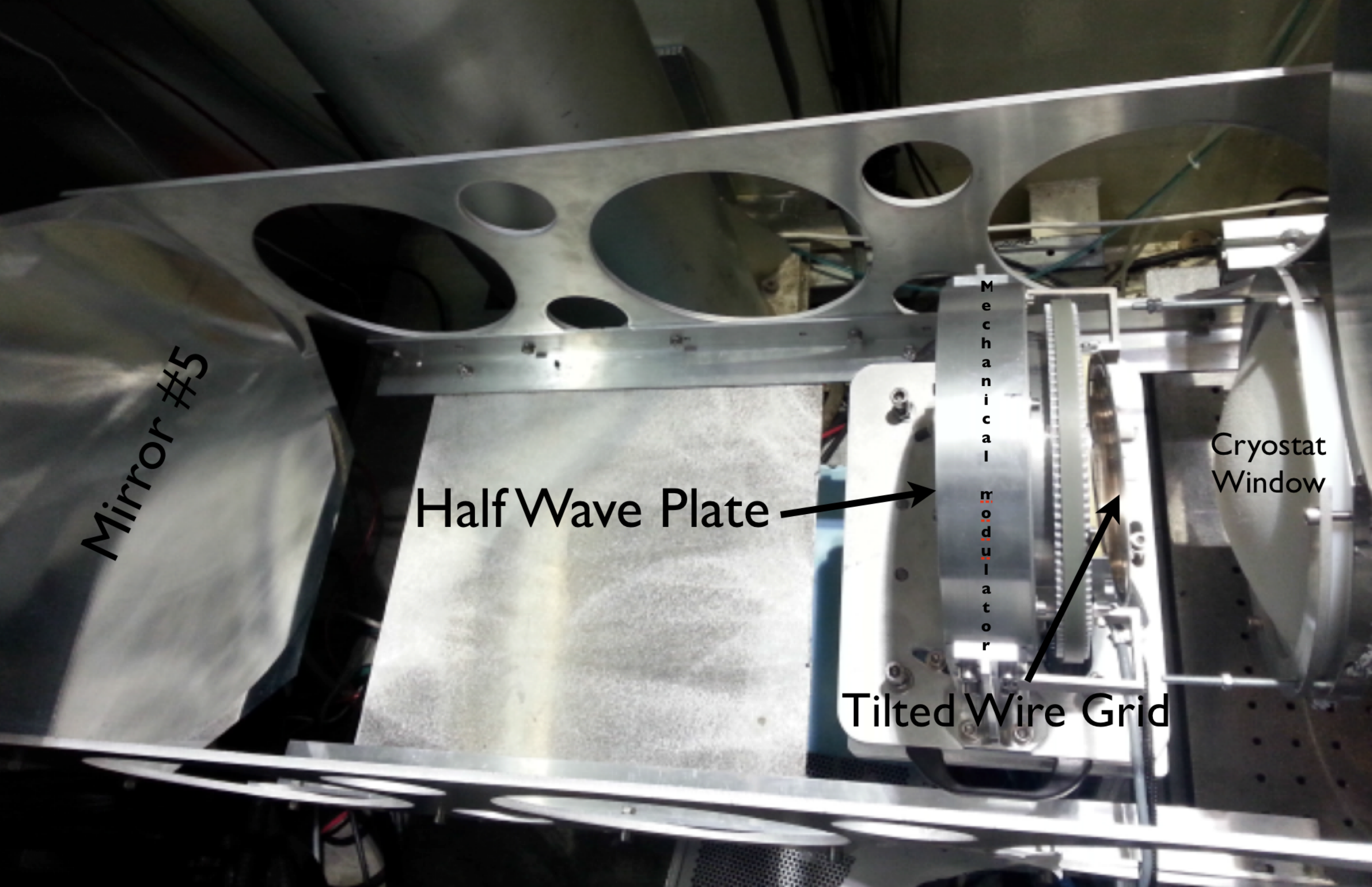}
\includegraphics[scale=0.25]{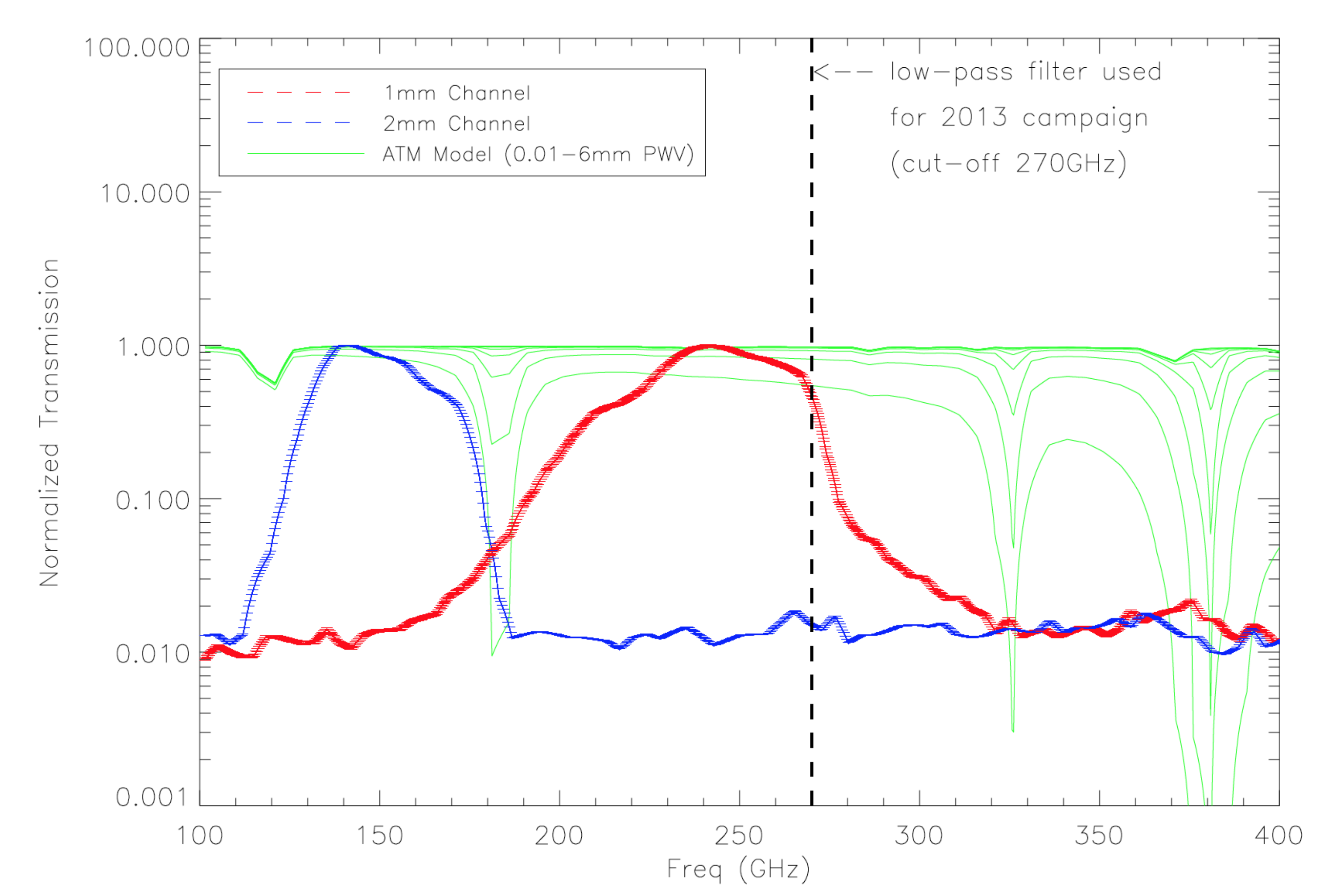} 
\end{center}
\caption{Top left panel: ray tracing from the entrance of the
  receiver cabin of the 30~m telescope (which is simulated but not shown on the
  image), view from the elevation axis (which is symbolized by the big circles
  and contain the 2 mirrors of the Nasmyth system). The NIKA cryostat and
  entrance nose are symbolised by the rectangles. To right panel: zoomed plot of the optical system inside the NIKA cryostat. Left bottom panel: NIKA polarisation facilities. Right bottom panel: NIKA bandpasses averaged over all valid pixels for 1.25~mm channel (red ) and 2.14~mm  channel (blue). The ATM model calculated for different water vapor contents is presented in green.}
\label{fig:optics}
   \end{figure}

\subsection{Optical coupling to the 30~m IRAM telescope}\label{ssec:optics}
The optical coupling between the telescope and the detectors consists in particular of a flat warm aluminium mirror at the top of the cryostat, an off-axis biconic-polynomial curved mirror, a 300~K window lens, a field stop, a 4~K lens, an aperture stop, a dichroic, a 100~mK lens and two band-defining filters in front of the back-illuminated KID arrays, which have a back-short matching the corresponding wavelength. In figure \ref{fig:optics} we present a view of the optics of the NIKA instrument (top panels) together with the coupling to the 30~m IRAM telescope. 
For polarisation measurements, we add the NIKA polarisation facilities that consists of a HWP is mounted at the pupil of the telescope (12~cm from the NIKA cryostat window) in a mechanical modulator performing the rotation. The LEKID design used in the NIKA prototype instrument are only sensitive to the intensity of the light, we put a polariser element (metal wire grid) at a distance of 6~cm from the HWP with its substrate plane at 10 degrees with the optical axis to avoid standing waves with the cold optical filters inside the cryostat. A picture  of the NIKA polarisation facilities is presented in figure \ref{fig:optics} (left bottom panel).

The shape, width and position of each NIKA band is achieved by a series of low-pass metal mesh filters, placed at different cryogenic stage in order to minimize the thermal loading of the instrument. In figure\ref{fig:optics} right bottom panel we present the NIKA band-passes together with a atmospheric model calculated for different water vapor contents. During the June 2013 observation campaign, an extra low-pass filter cutting frequencies above 270~GHz (see vertical line in figure \ref{fig:optics} right bottom panel) was added since it was suspected that the optical load on the 1.25~mm array was preventing it from reaching the appropriate operating temperature. In the final configuration this filter was removed. 


\subsection{The detectors and the readout electronics}\label{ssec:detectors}

The adopted solution for the NIKA instrument uses dual polarization Lumped Element Kinetic Inductance Detectors (LEKIDs) using a Hilbert pattern inductor\cite{vinoy}. To fully exploit the dual-polarization, spectral band splitting is achieved using a dichroic beam splitter\cite{Roesch2012}. Over several observation campaigns between 2012 and 2014, various KIDs arrays have been tested on each of the two focal planes with the final array designs now fixed as follows:
\\
\\
\\
\textbf{2.14~mm channel array:} the array is made of 132 pixels obtained from an 18~nm Aluminium film on a 300~$\mu$m HR Silicon substrate. Cross-bonding the ground-planes of the co-planar feed-line has been added to suppress spurious slot-line modes that were affecting the uniformity, leading to an increase in the number of valid pixels. 
\\    
\\
\textbf{1.25~mm channel array:} The 1.25~mm array is made of 224~pixels with a an 18 nm Aluminium film on a 180~$\mu$m substrate. The number of valid pixels during this campaign 125 at 2.14~mm and 190 at 1.25~mm, for a total of 315 pixels.
\\
\\
\\
The readout comprises of the new NIKEL electronics described in {\emph Bourrion et al.}\cite{Bourrion2012}. We describe briefly the idea:  6 separate Field Programmable Gate Arrays (FPGAs), are coupled to a Digital to Analog Converter; they can generate a comb of frequencies (up to 400 tones) each over a 500~MHz bandwidth each set to the resonant frequency of a LEKID in the array. The comb is then up converted by mixing with a local oscillator carrier at the appropriate frequency. The resulting signal is fed into the cryostat using a single coaxial line, and excites each resonator. On the output line, one cryogenic low noise amplifier boosts the signal of each read out tone, which then is down-converted to the base band and is acquired by an Analog to Digital Converter (ADC). The output tones are compared to a copy of the input tones kept as a reference, so that for each pixel it is possible to measure the component of the signal that is in phase (I) and in quadrature (Q) with respect to the reference. Five of the FPGAs generate 80 tones each over a 100~MHz band, using five associated DACs. The sixth FPGA acts as a central unit that combines the signal of the other units, appropriately shifting and filtering the different 100~MHz sub-bands to finally cover the whole 500~MHz available for the frequency comb used to excite the detectors. An analogous,
but reversed, process is then applied to the signal acquired by the ADC of the board, which is once again split in 5 different sub-bands treated separately.

\newpage

\section{The whole calibration process}\label{sec:cal}
The variety of NIKA scientific targets range from thermal SZ observations to
dust polarization properties. This wide range of applications requires, a controlled readout technique to properly track the change in resonant frequency of the detectors, an accurate photometric calibration process able to define as well as remove the impact of systematic errors on the final sky maps and a dedicated data reduction pipeline. The calibration process has been described in \emph{Catalano et al.}\cite{NIKA2014}. We list in the following the critical points to be controlled:  

\begin{itemize}

\item {\bfseries Readout optimisation technique:} One of the most difficult challenges in operating with KIDs, is to convert the observed in phase (I(t)) and in quadrature (Q(t)) signal to absorbed optical power. This is a very different task with respect to the dissipative readout like thermal detectors (high impedance bolometer, TES, etc...). One possible solution is to perform a frequency sweep before starting each scan in the sky and to determine the centre of the resonance circle (I,Q) and calibrate the change of the phase ($\phi(f)$) as a function of frequency. The validity of this method depends directly on the stability of the atmosphere. Indeed, if the sky emission fluctuates during a scan, the resonance circle changes and therefore the responsivity of the detector changes. In order to improve the photometric reproducibility we developed a system to control the change of the signal by modulating the frequency on the local oscillator (by few kHz) synchronously to the FPGA in order to generate two tones one just below the resonant frequency of the detector. An average of 50 points is then performed in the FPGA acquiring data at the rate of 23.842 Hz. With this method it is possible to estimate the variation of the resonant frequency of the detectors $\Delta f_0(t)$, by projecting $(\Delta I(t), \Delta Q(t))$ along the gradient found as:
\begin{equation}
\Delta \hat{f_{0}}(t) = \frac{\left(\Delta I(t), \Delta Q(t)\right)\cdot\left(dI/df(t), dQ/df(t)\right)}{\left(dI/df(t), dQ/df(t)\right)^2}\cdot\delta f_{LO}
\label{eq:RFdIdQ}
\end{equation}

A detailed description of this method is given in \emph{Calvo et. al}\cite{Calvo2013} and \emph{Catalano et al}\cite{NIKA2014}.

Another improvement performed during the previous observation campaigns was an automated tuning procedure.  
When operating from ground, the variations in the background load due to the atmosphere can cause the resonant frequency of the detectors to vary by a substantial amount, introducing shifts that can be in some cases larger than the resonances themselves. This effect must be constantly monitored and, if needed, counterbalanced by changing the excitation tones, in order to always match the resonant frequency of each pixel and to thus probe each detector near to its ideal working point. The new tuning method is based on the measurement of the angle $\varphi$ between the vectors ($I$, $Q$) and ($dI/df_{LO}$, $dQ/df_{LO}$). Thus, a single data point is now sufficient to retune the detectors, without recurring to frequency sweeps. This leads to a crucial advantage in terms of observing time, especially in the case of medium and poor weather conditions. The new procedure reduces the required time for retuning by 75~\%. A detailed description of this method is given in \emph{Catalano et al.}\cite{NIKA2014}.

\item {\bfseries Field Of View (FOV) reconstruction : } In order to recover the pointing direction of each pixel, a focal plane reconstruction via planet scans is used. We scan a bright astronomical source at 35~arcsec/s with the entire focal plane. The astronomical calibration source is a planet (Mars, Uranus, Neptune), which has a small angular diameter compared to our beam and can be considered as a point source for most of the NIKA observations. The position, width and the orientation of each KID are determined by building a template of the low frequency part of the signal (mostly sky and electronic noise) using all detectors that are far from the source (typically more then 30~arcsec). More details are presented in \emph{Catalano et al.}\cite{NIKA2014} and \emph{Adam et al}\cite{NIKA_SZ}.
The accuracy of this technique is presented in table~\ref{table:perf}.

\item {\bfseries Photometric calibration:}  Photometric calibration defines as well as possible the impact of systematic source of errors in the final sky maps. These sources of systematic errors come from \emph{external} uncertainties (calibration factors deduced from primary calibrator sources) or internal instrument uncertainties such as spectral response uncertainty, secondary beam fraction and opacity correction. In particular the opacity correction procedure was improved during the last observation campaign. To derive the opacity at the exact position of the scan and at the same frequencies as NIKA, we have implemented a procedure in order to use the NIKA instrument itself as a tau-meter measuring the variation of the resonance frequencies of the detectors versus the airmass via elevation scans (\emph{sky-dips} \cite{dicke}). The procedure is based on the idea that the KIDs response (the change in resonant frequency for a given 
change in absorbed power) is a constant property for each detector. This has been demonstrated in laboratory under realistic conditions (with a optical load changing  between about 50~K and 300~K, \cite{monfardiniLTD}).
With this method we can derive coefficients that depend only on the response of the detectors  to be applied in the selected map to calculate the opacity in the considered scan. With this procedure we do not need to perform the sky-dips at the exact time of each source observations and therefore we can estimate the atmospheric opacity at the same position (Azimuth-Elevation) of the source, instead of using the average sky opacity. 

We estimated that the overall calibration uncertainty for point sources on the final data at the map level is around 15~\% for 1.25~mm channel and 10~\% for 2.14~mm channel (see table \ref{table:perf}). This is based on the observed rms flux variation on a
calibration source\cite{NIKA2014}. 

\item {\bfseries Data Reduction Filtering : } 
We developed a dedicated reduction pipeline to calibrate, filter and process data onto sky maps. The principal steps of the processing are the raw data creation (raw data and the main instrumental parameters), the flagging of bad detectors depending on the statistical noise properties, data filtering (dedicated Fourier space filtering removing of frequency lines produced by the vibrations of the pulse tube), absolute calibration procedure (calibration factor deduced from primary calibrator and atmospheric absorption correction), Atmospheric and electronic noise decorrelation (different procedure depending on the scientific target to be reached) and map making.
The estimated systematic error introduced by the data reduction filtering (in particular due to the common modes subtraction) is estimated to of order 5~\% for both channels in the case of point sources. This was computed from the dispersion between different
processing modes.

\end{itemize}

In table \ref{table:perf} we present the performance of the NIKA instrument updated at the last technical observing campaigns performed in June 2013. The Noise Equivalent Flux Density (NEFD) is calculated as the array averaged sensitivity to point sources and corresponds to the rms flux obtained in one second of integration. We have obtained a sensitivity (averaged over all valid detectors) of 40 and 14~mJy.s$^{1/2}$ for best weather conditions for the 1.25~mm and 2.14~mm arrays respectively. This level of sensitivity is very close to the photon noise at the 30~m IRAM site. Additionally, the camera performance can be quantified with its mapping speed: the area that can be observed per unit time at a given sensitivity.

\begin{table}[h]
\caption{Summary of the NIKA characteristics and performance. $^{*}$The sensitivities presented correspond to the ones estimated during the technical observing campaign of June 2013.}
\label{table:perf}
\begin{center}       
\begin{tabular}{|l|l|l|} 
\hline
{\bfseries Array} & 1.25~mm & 2.14~mm  \\
\hline
{\bfseries Valid Pixels} & 190 & 125 \\
\hline
{\bfseries Field Of View}~[arcmin] & 2.2 & 2.2 \\
\hline
{\bfseries Band-Pass}~[GHz] & 200-280   & 125-175  \\
\hline
{\bfseries FWHM}~[arcsec] &  12.3  & 18.1 \\
\hline
{\bfseries Overall Calibration Error}~[\%] &  15  & 10 \\
\hline
{\bfseries Sensitivity}~[mJy$\cdot s^\frac{1}{2}$]$^{*}$ & 40 & 14  \\
\hline
{\bfseries Mapping Speed}~[arcmin$^2$/mJy$^2$/hour] & 8 & 57  \\
\hline
\end{tabular}
\end{center}
\end{table}

\section{NIKA commissioning: observation campaigns 2012-2014}\label{sec:run5-6-7}

The NIKA camera has been used during the several technical observing campaigns (November 2012, June 2013, January 2014) to observe point-like sources, in order to assess the NIKA photometry and extended sources to demonstrate the possibility to reconstruct angular scales up to few arcminutes. Finally we tested the NIKA polarisation capability in January 2014 observing campaign. In the next subsections we present some relevant result of these observing campaigns. 

\subsection{Extended sources}

\begin{figure*}[t!]
\begin{center}
\includegraphics[scale=0.19]{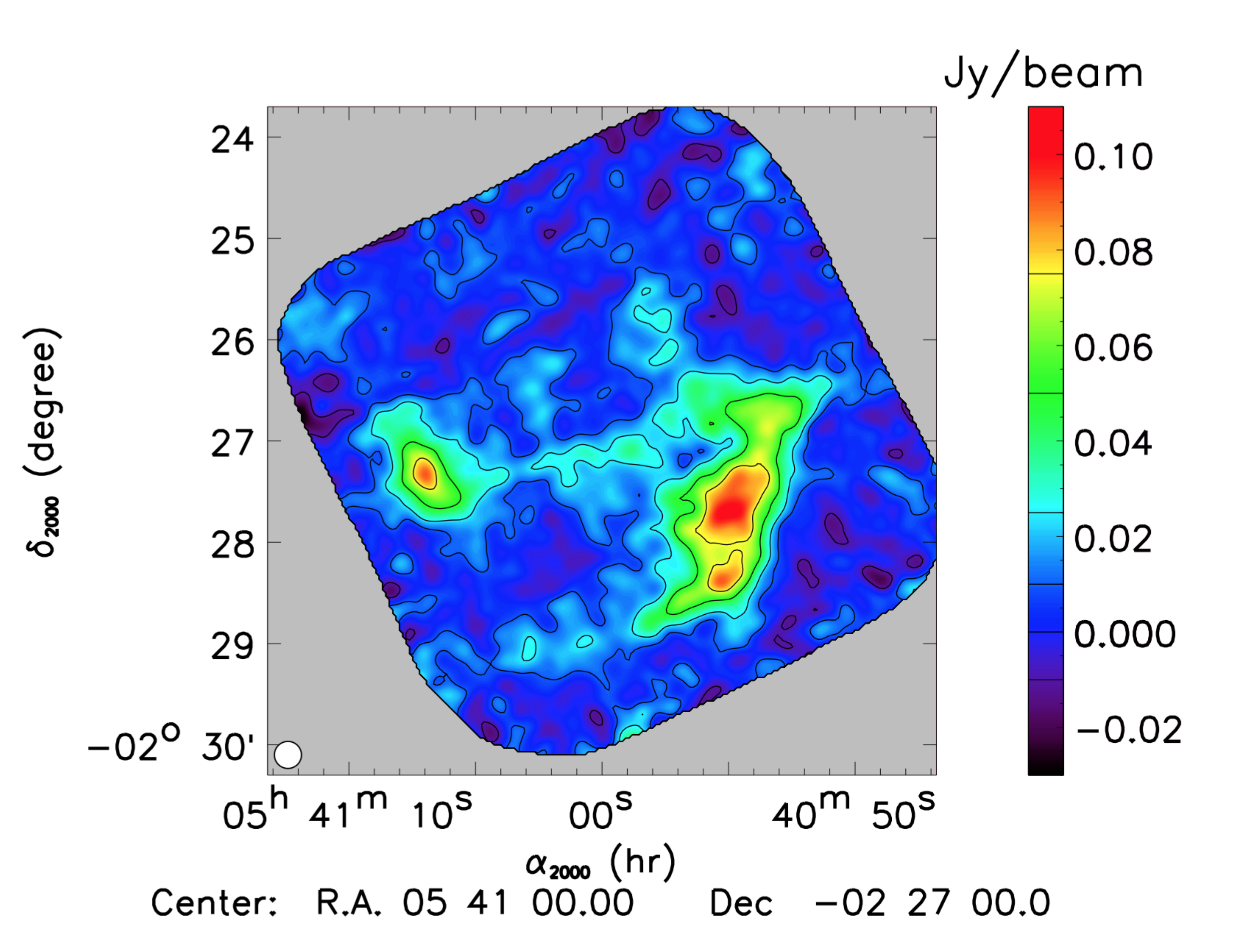} 
\includegraphics[scale=0.19]{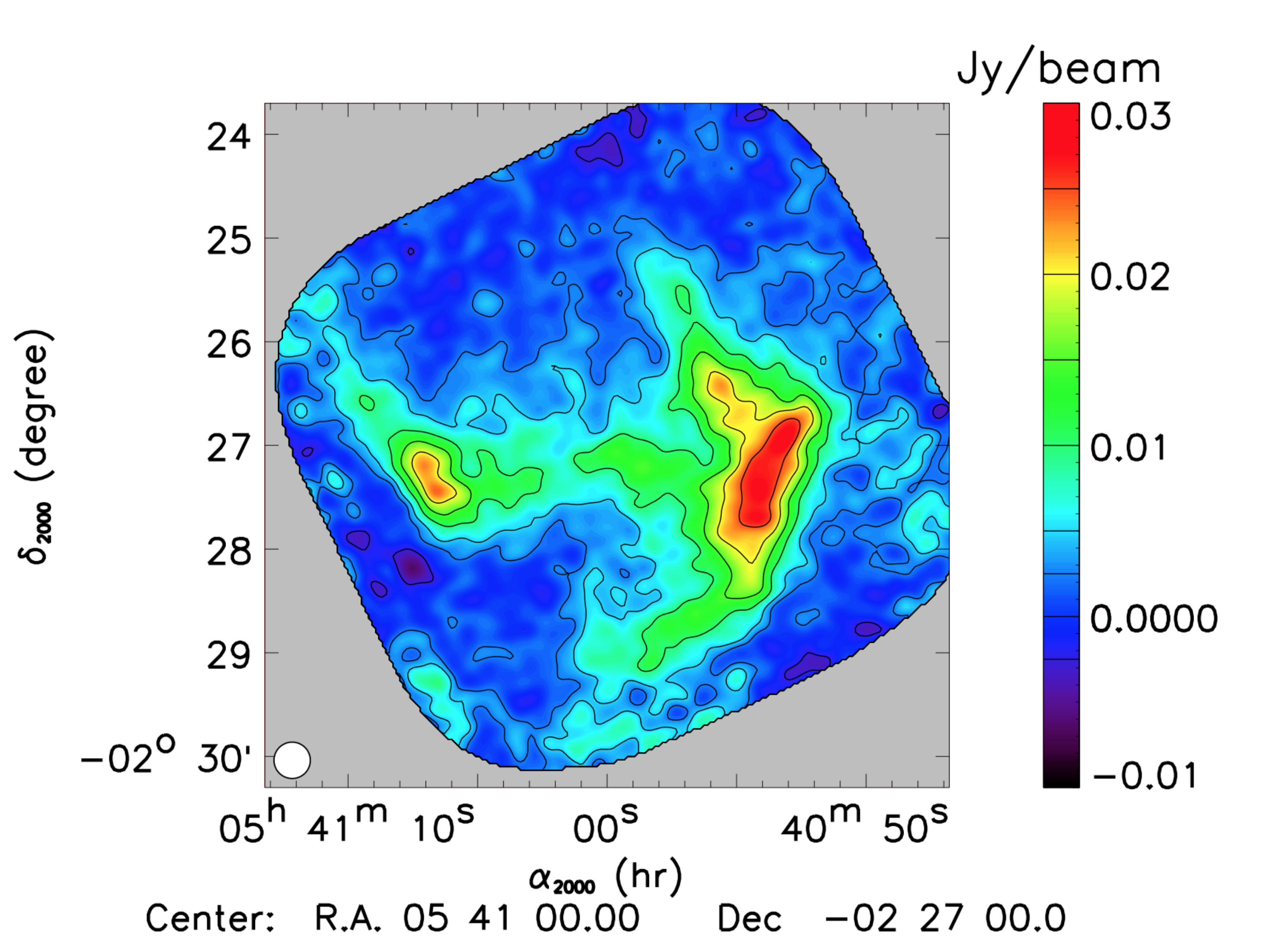}
\includegraphics[scale=0.19]{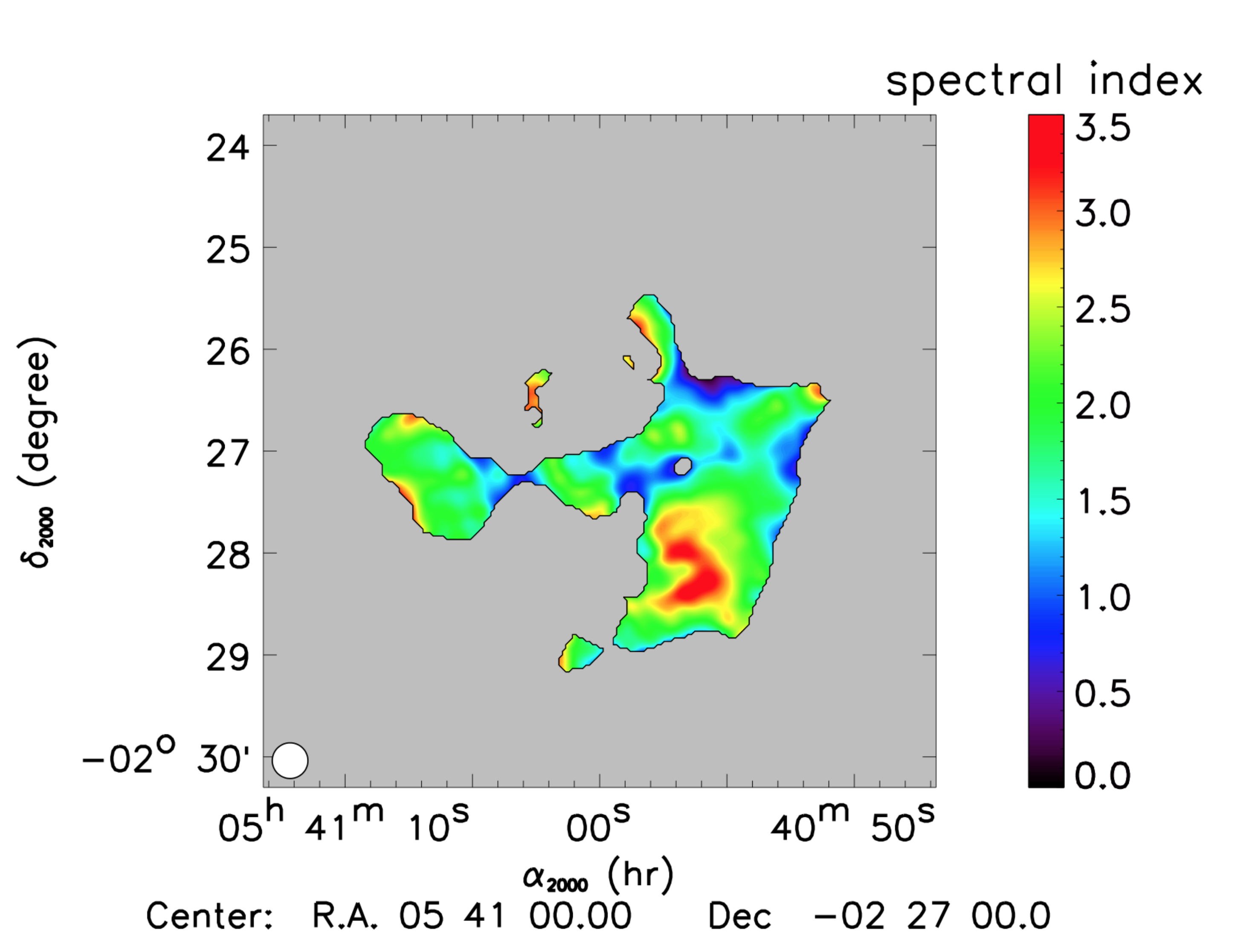}
\end{center}
  \caption{Examples of 1.25 (left) and 2.14~mm (middle) NIKA maps of  Horsehead nebula. Spectral index maps are presented on the right column. Observing date: 2012-11-23. Integration time: 1.57 hours.}
\label{fig:extended_sources}
   \end{figure*} 

During the 2012 and 2013 observation campaigns we have observed simultaneously
at 1.25 and 2.14~mm several well-known extended sources to test the capabilities of
the NIKA camera to recover large angular scales up to few arcminutes. We performed
both elevation and azimuth on-the-fly zig-zag scans to ensure a homogeneous coverage of
the mapped area. The size of the scans and integration time have been adapted
to each source. We present an example of extended source, namely the
Horsehead Nebula observed during the 2012 observation campaign\cite{NIKA2014}. 

The NIKA 1.25~mm is consistent with the
1.2\,mm continuum map \cite{2005A&A...440..909H} obtained with MAMBO2, the
MPIfR 117-channel bolometer array \cite{1992ESASP.356..207K}.  We clearly
observe in the NIKA maps the photon dominated region whose morphology changes significantly
from one frequency to another. This is also obvious in the spectral index map
presented in the right figure that ranges between 2 and 5. 
The northern part of the PDR presents a
significantly flatter spectral index (about 2) than the southern part and
the main (about 3.5). Two possible explanations are: CO 2-1 contamination at 1
mm and/or high dust emissivity spectral index\cite{2013ApJ}. 
There is also evidence of 2~mm emission that may be due to free-free emission.

\subsection{Sunyaev-Zel'dovich observations}

	\begin{figure}[b!]	
	\centering	
	\includegraphics[scale=0.31]{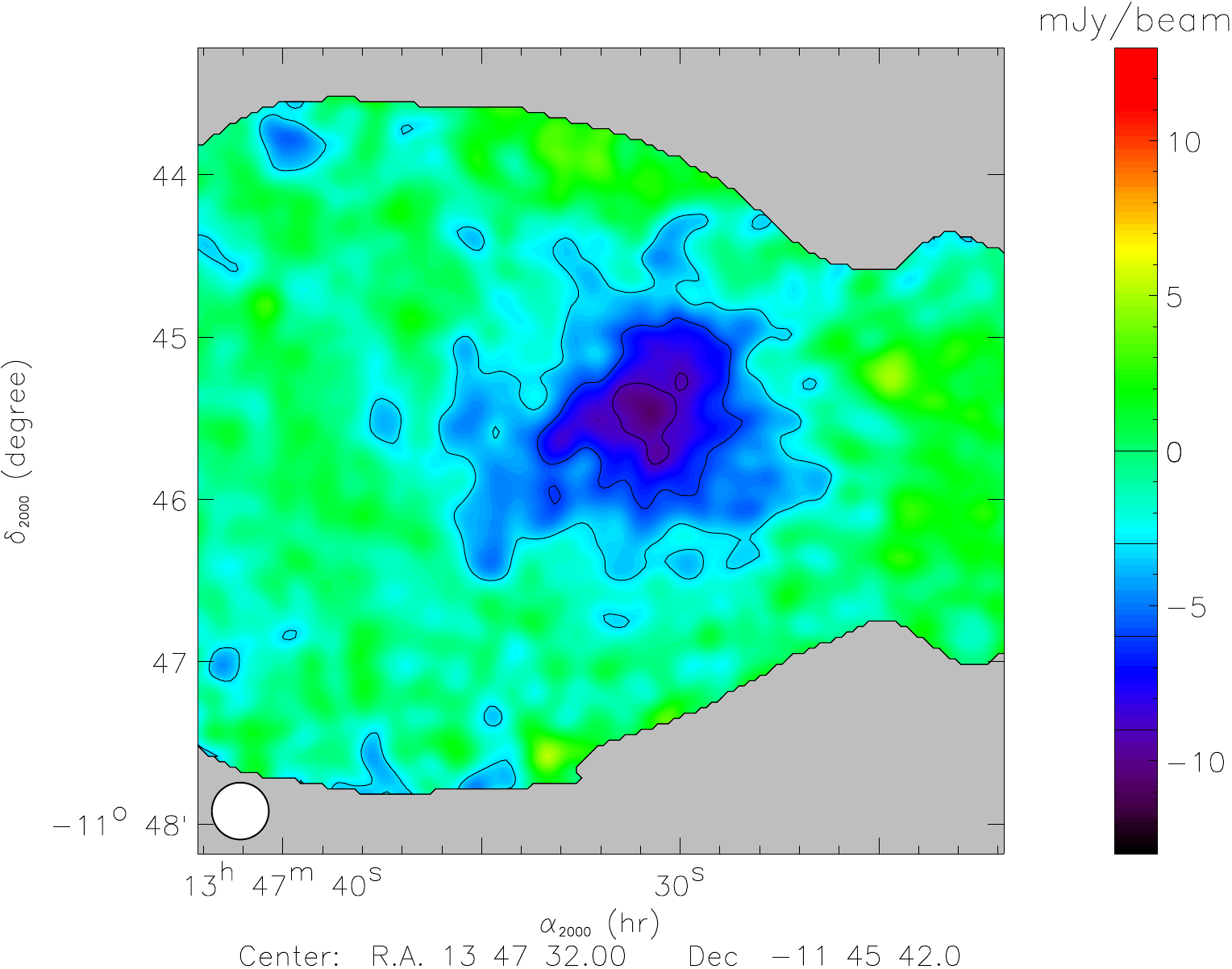}
		\hspace*{0.6cm	}
	\includegraphics[scale=0.31]{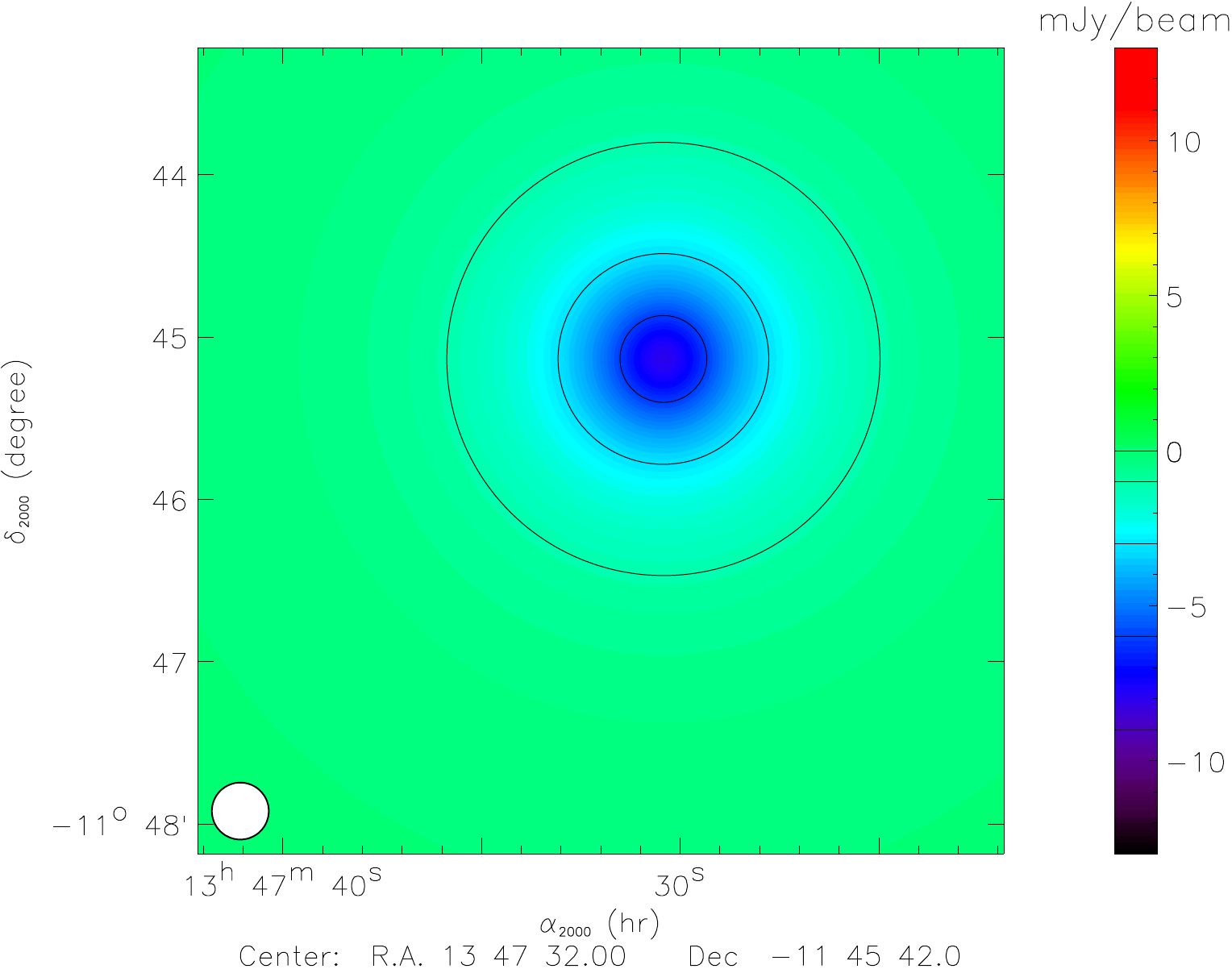}
		\hspace*{0.6cm}
	\includegraphics[scale=0.31]{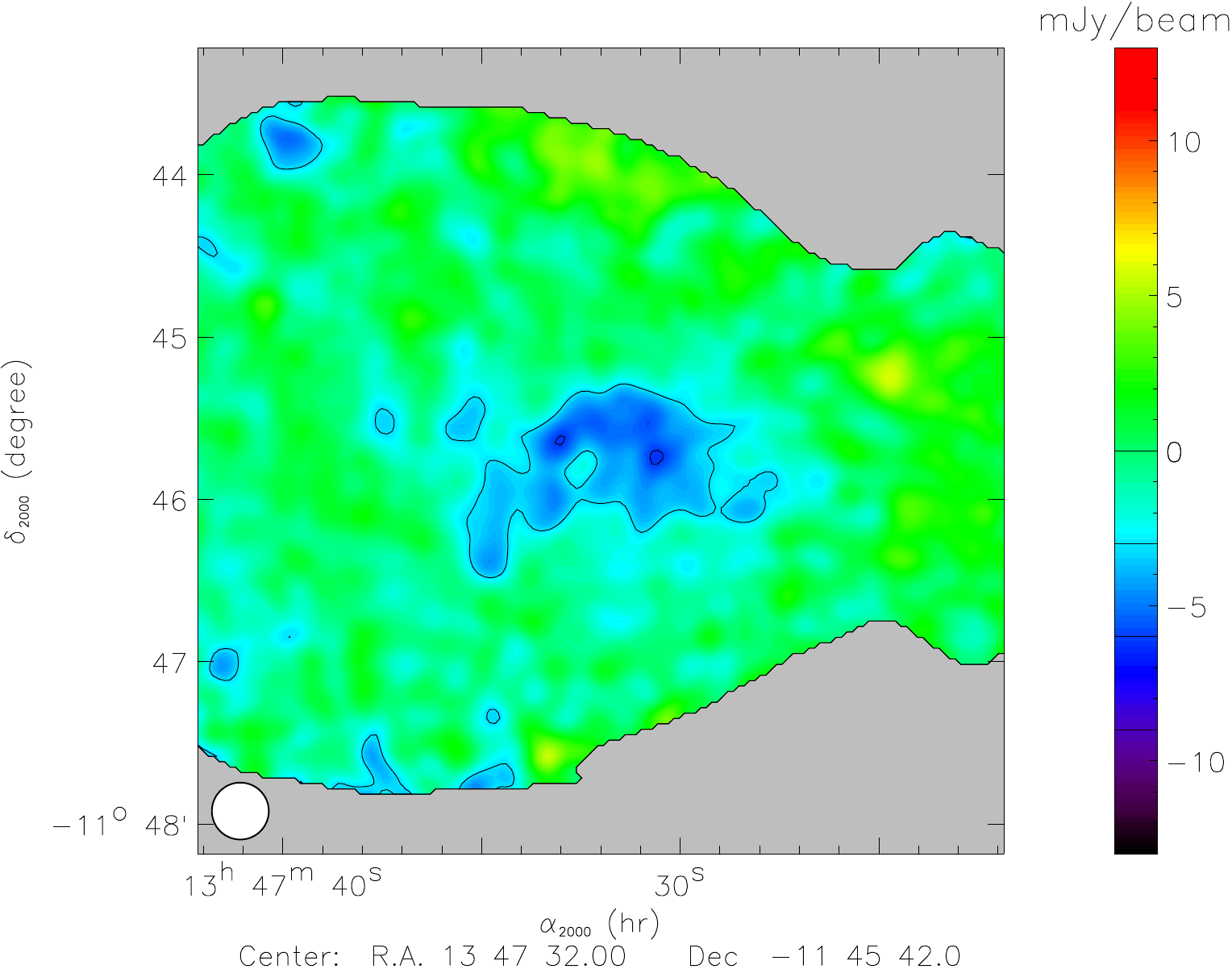}
	\caption{Comparison between the original point source subtracted RX~J1347.5-1145 tSZ map (left panel) and the best fit model map excluding the shock area (middle panel). The residuals are given on the right panel map. The model accounts well for the cluster emission except in the southern shocked area as expected. Observing dates: 2012-11-21/22/23. Total integration time: 8 hours.}
        \label{fig:residual}
	\end{figure}

Technical observing campaign during 2012 has proved that KID arrays are competitive detectors for the observation of galaxy clusters via the Sunyaev-Zeldovich effect. The thermal Sunyaev-Zel'dovich (tSZ) effect\cite{SunyaevZeldovich} is a powerful cosmological probe and a way to understand the complex gravitational and non-gravitational processes acting in galaxy clusters. In particular the NIKA team chose RX~J1347.5-1145\cite{NIKA_SZ} cluster as a good target for the first SZ observations with NIKA instrument. We imaged the tSZ morphology of the cluster on reliable scales between 20 and 200~arcsec from the core to its outer region using a dual-band de-correlation. Results have been validated using simulation and show a strong South East extension that corresponds to the merger shock also observed in the radial flux profile of the cluster and the residual of the map with respect to the modelling of the relaxed part of the cluster \cite{NIKA_SZ}. This strong extension was expected from the overpressure caused by the ongoing merger\cite{point1,point2,koma}.

\subsection{NIKA Polarisation first light}

 \begin{figure}[t!]
\begin{center}
\includegraphics[scale=0.15]{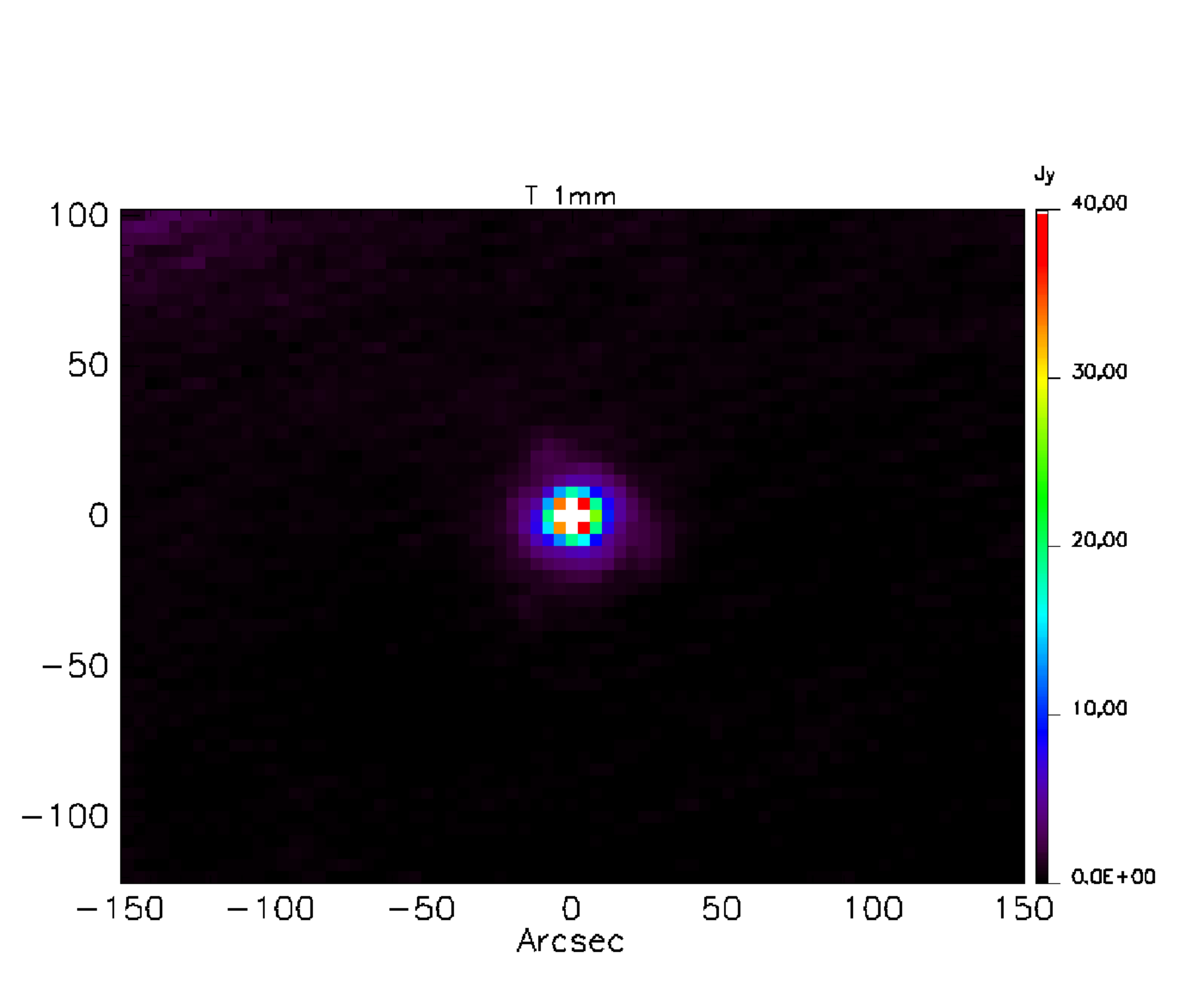}
\includegraphics[scale=0.15]{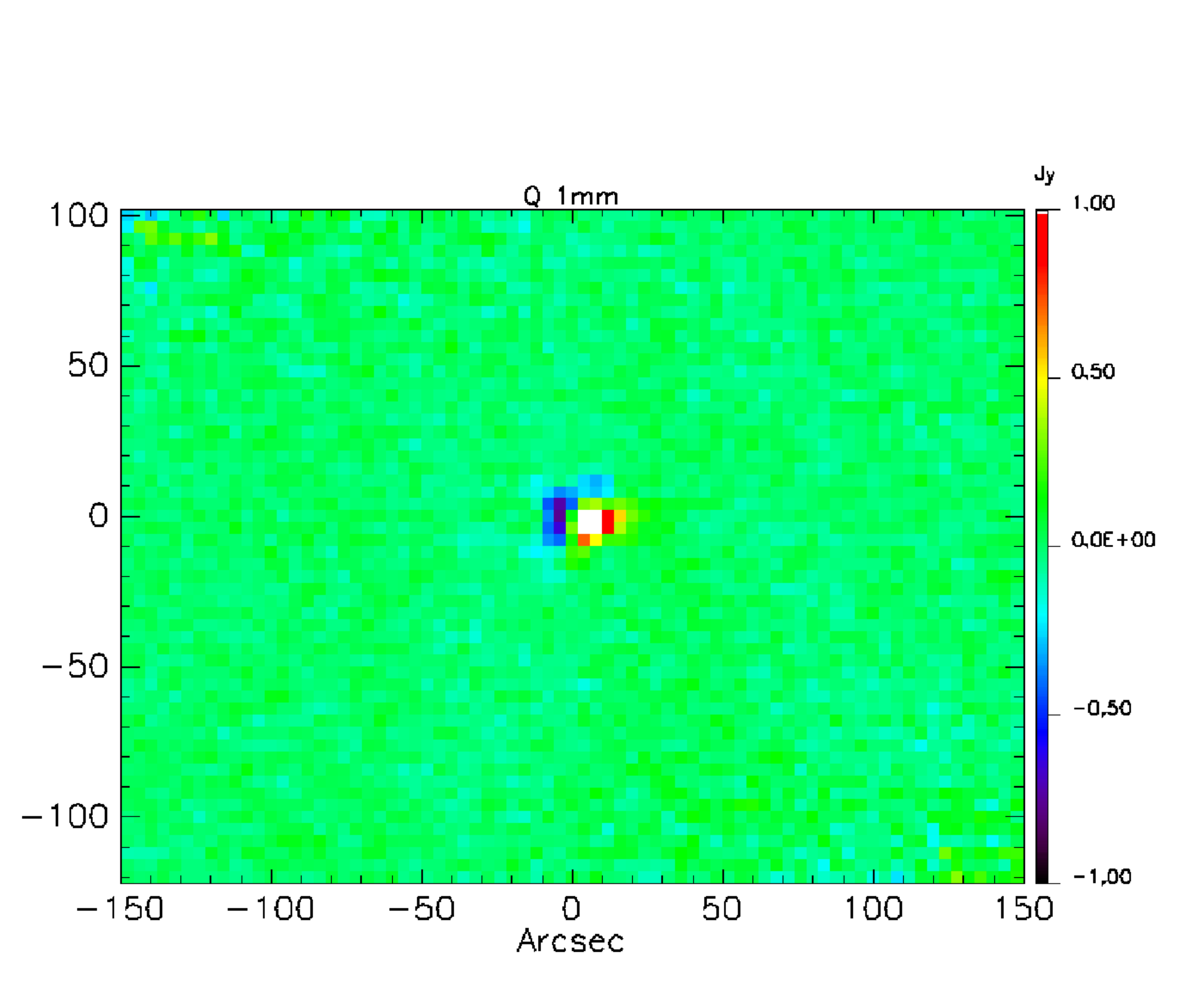}
\includegraphics[scale=0.15]{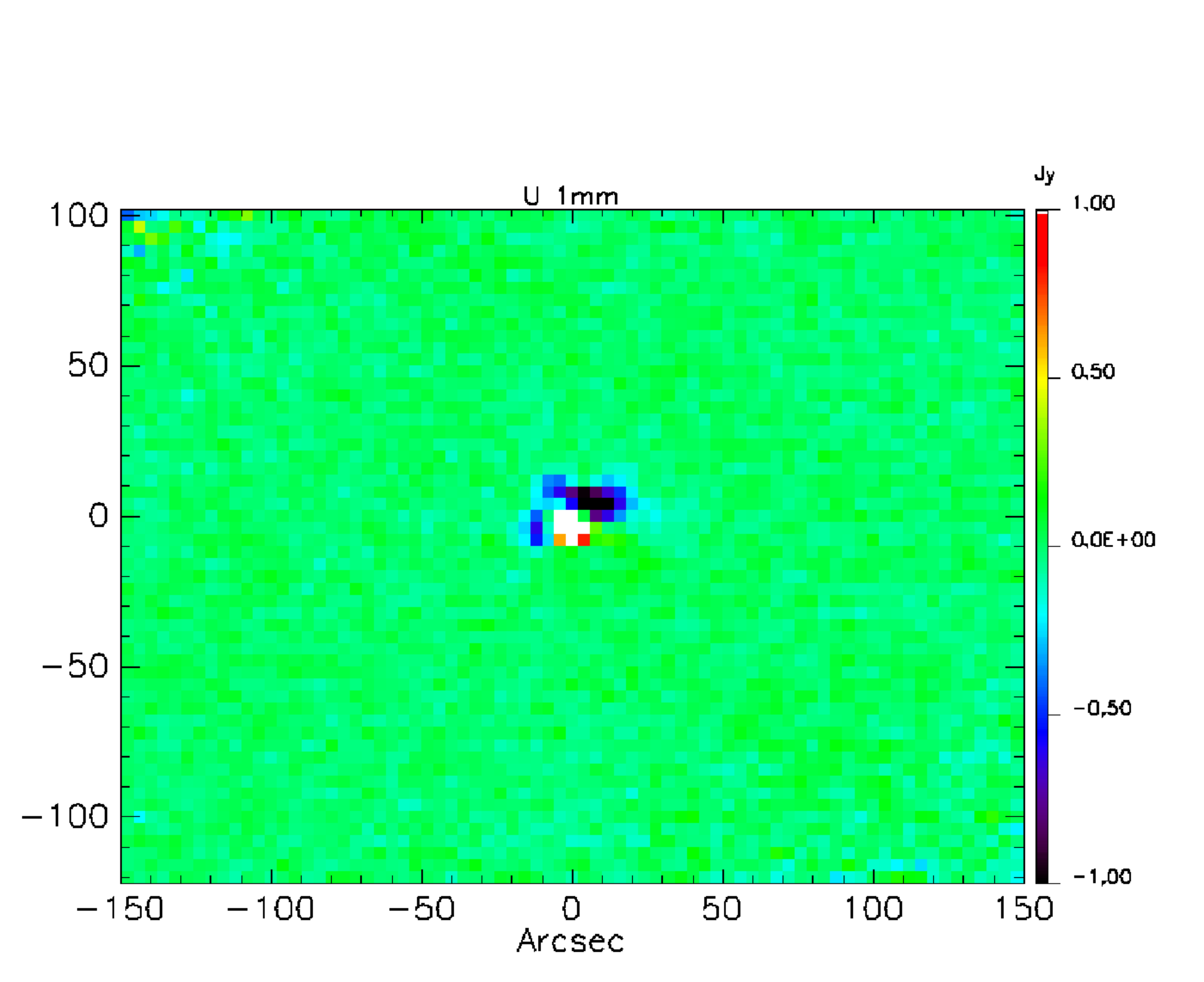}
\end{center}
\caption{Instrumental polarisation characterisation using Uranus polarized map: the level of intrumental polarization is at the level of 2-3\%.}
\label{fig:uranus_pol}
   \end{figure}

The NIKA polarisation facility comprises a rotating HWP located in the pupil of the telescope and a 300~K wire grid polariser mounted at a distance of 6cm from the HWP with its substrate plane at 10 degrees as shown in figure \ref{fig:optics} (left bottom panel).
This setup permits simultaneous measurements of the three Stokes parameters (I, Q, U) on
a same area of sky through the same optical path in the telescope. We chose the continuous rotation strategy to modulate the
polarisation. In this mode, a linearly polarised signal is modulated by the combined action of an
\emph{ideal} HWP and subsequent analyser at four times the mechanical rotation frequency. It can
hence be extracted with a lock-in procedure by isolating the amplitude of the fourth harmonic
of the mechanical rotation where the modulation function of the polarised signal is located. The
accuracy of this strategy lies in the calibration of the HWP: the correct amplitude of the various
Fourier components relative to different input signals is required in order to decouple unwanted
instrumental contaminations from the desired signal.
The performance of the polarisation modulation has been tested during the NIKA technical observing campaign in January 2014. 
The chosen design of the HWP for the first run of measurement is a single bi-refringent sapphire
plate. This kind of mono-chromatic HWP was preferred to a more sophisticated achromatic
HWP to minimise and control the polarisation of systematics for future NIKA and NIKA2 polarisation measurements. 
The plate is coated on both sides with a single dielectric substrate in order to minimise
surface reflections at the mean wavelength of the known spectral band (230-
240~GHz). In figure \ref{fig:uranus_pol}, we present preliminary I, Q and U maps of Uranus which is supposed to be unpolarised, observed for the 1.25~mm channel. The non zero signal seen in Q and U is a hint of instrumental polarization at the level of the order of 2-3~\%.

 \begin{figure}[t!]
\begin{center}
\includegraphics[scale=0.35]{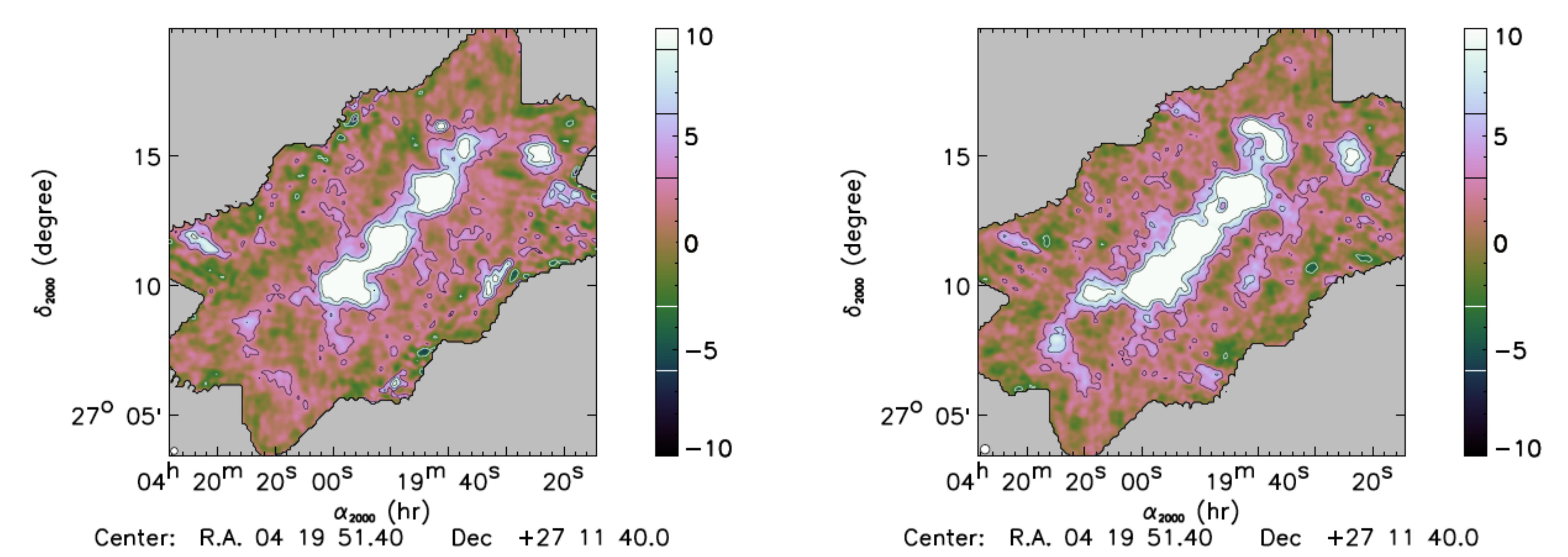}
\end{center}
\caption{Taurus main filament first results from february 2014 NIKA open pool: 1.25~mm channel (left) and 2.14~mm channel (right) signal to noise ratio.}
\label{fig:taurus}
   \end{figure}

\section{The first scientific open pool}\label{sec:run8}

\begin{figure}[b!]
\begin{center}
\includegraphics[scale=0.24]{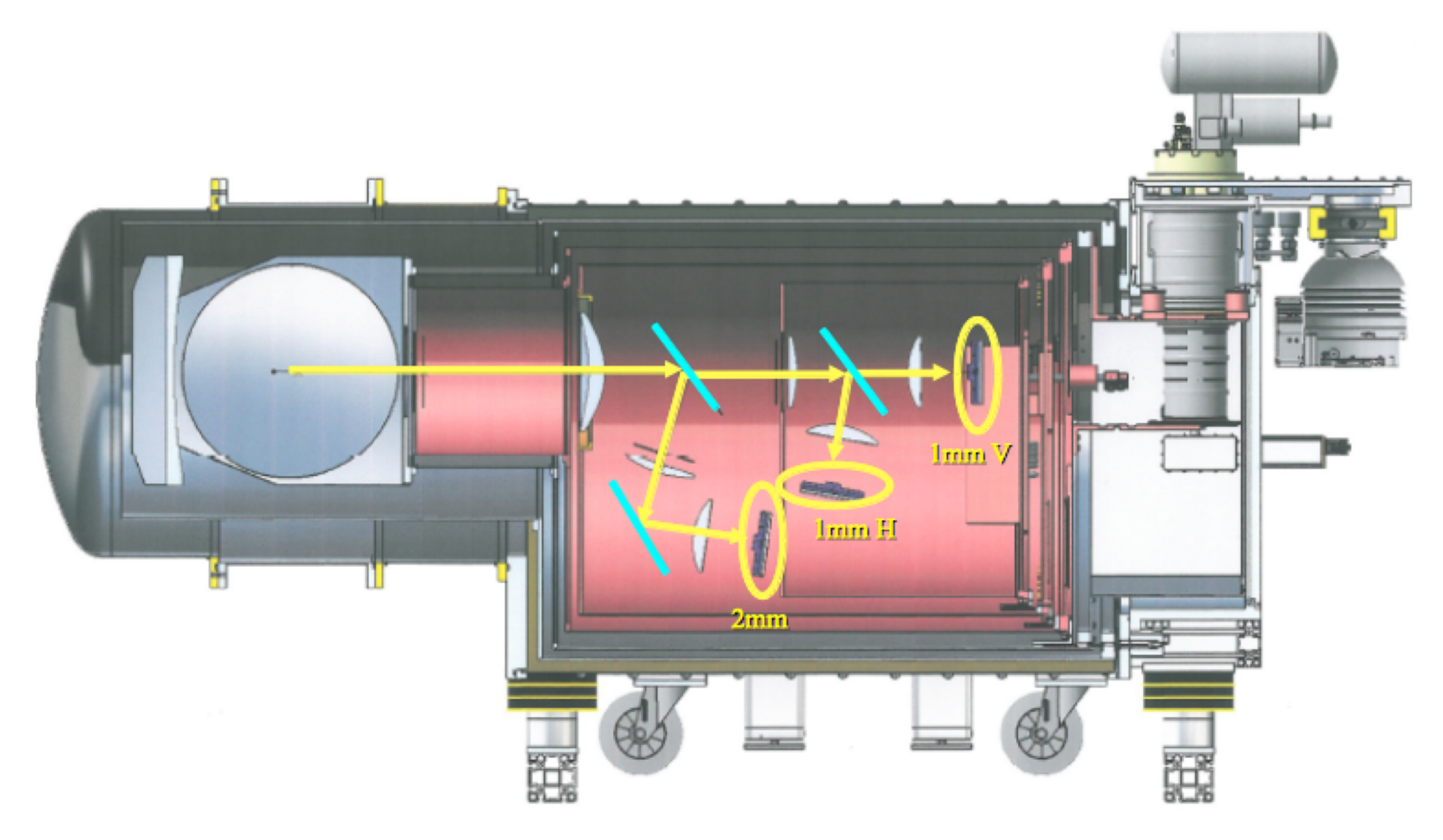}
\includegraphics[scale=0.24]{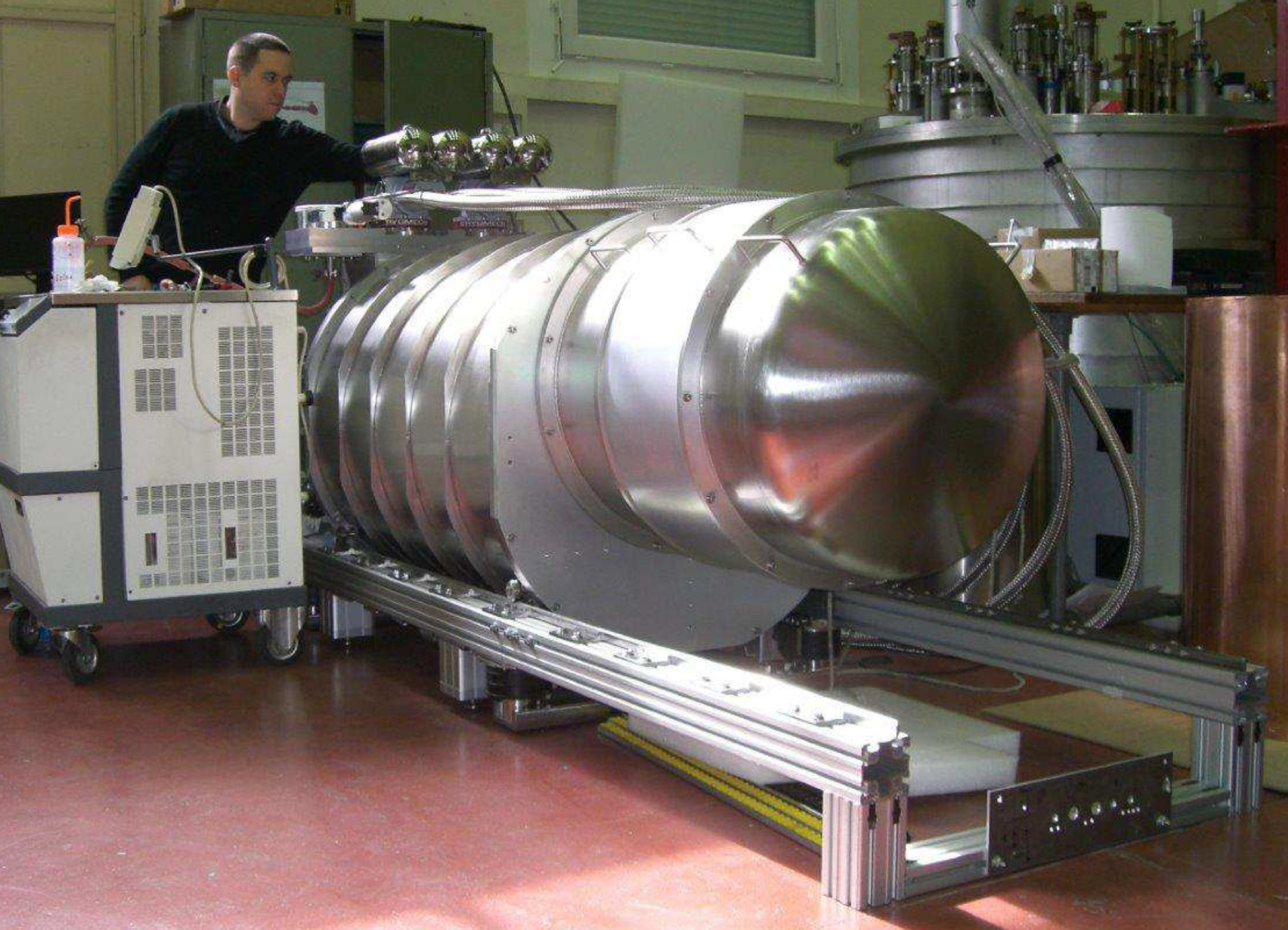} 
\end{center}
\caption{NIKA-2 instrument conceptual draw (top panel) and picture (bottom panel).  The beam splitting is achieved by a dichroic and a polariser. The three KID arrays are cooled down at roughly 100~mK. Total length: 2.3~m; weight: 980~kg.}
\label{fig:nika2}
   \end{figure}

In February 2014 the NIKA instrument performed the first open pool. It comprised 162 hours of observations with 108 hours of open time (for astronomers outside the NIKA collaboration) and 54 hours of guaranteed time (restricted to the NIKA collaboration). Thirteen open time proposals have been accepted by the IRAM program committee. These proposals cover a large variety of scientific targets from the solar system science to the $z \ge 4$ universe. The data products which the NIKA collaboration provides to external observers for the open time observations are the calibrated Time Ordered Data, the maps, calibration results (measured beam, focal plane geometry) and log files.

The NIKA guaranteed time proposals comprised of three themes with of 18 hours of observing time each:

\begin{itemize}

\item {\bfseries tSZ mapping of high redshift galaxy clusters : } observations of two of non-relaxed intermediate and relaxed high redshift galaxy clusters (MACS J0717.5+3745 at z=0.55 and CL 1226.9+3332 at z=0.89 respectively) in order to improve our understanding of the physics of the substructure level. 

\item {\bfseries Confirming $z \ge 2$ proto-cluster candidates observed by Planck and Herschel : } this follow-up program intends to observe 23 massive high redshift galaxy clusters to reconstruct the Spectral Energy Distribution (SED) and dust properties and ultimately to measure the photometric redshift including also GISMO and SCUBA2 results.  

\item {\bfseries Probing the inner structure of the Taurus main filament : }  observations of a segment of the Taurus main filament with the dual-band capability of NIKA. Combining NIKA results with Herschel $70-250 \mu m$ data we can reconstruct maps of dust temperature, column density and dust emissivity. In figure \ref{fig:taurus} we present a signal to noise map for each NIKA channel as a very early result of this study.

\end{itemize}

\section{From NIKA to NIKA2}\label{sec:nika2}
The next generation NIKA2 camera will cover a field-of-view of 6.5~arcmin larger then the 2~arcmin of NIKA. It will keep the dual-band imaging capabilities and, at the same time it will measure the linear polarisation
on the 1.25~mm channel. This will be achieved by using a dichroic beam splitter to define each spectral band with the 1.25~mm band being split into two polarisations using a polarizing beam splitter. Modulation of the polarization is then achieved with a rotating half wave plate. The resulting NIKA2 focal planes will therefore consist of three large LEKID arrays (sensitive area diameter = 80~mm) and a total pixels count of 5000. The first prototype of the NIKA2 array (1020 pixels at 150~GHz) has been produced and tested in Grenoble. The readout solution for the NIKA2 instrument is the same NIKEL system described previously. 
A picture and a mechanical draw of the NIKA 2 instrument is presented in figure \ref{fig:nika2}.  
NIKA2 is fabricated in Grenoble, is officially selected by IRAM as the next generation continuum and polarisation instrument at the 30~m telescope. It will be installed for Commissioning in 2015.

\section{Conclusion} 
The NIKA instrument is now open to the 30~m IRAM telescope observers from the Winter 2014. From technical and scientific observing campaigns we demonstrated competitive sensitivities at both 1.25 and 2.14~mm and good photometry performances. In addition, during the first scientific open pool we acquired good quality data and work is still in progress to produce promising results. In 2015 the NIKA instrument will be replaced by the next generation dual-band NIKA2 camera 
for continuum and polarisation measurements using a total of 5000 pixels shared between three focal planes at 1.25~mm and 2.14~mm. 

\small
\section*{ACKNOWLEDGMENTS}       
 We would like to thank the IRAM staff for their support during the campaign. 
This work has been partially  funded by the Foundation Nanoscience Grenoble, the ANR under the contracts "MKIDS" and "NIKA". 
This work has been partially supported by the LabEx FOCUS ANR-11-LABX-0013. 
This work has benefited from the support of the European Research Council Advanced Grant ORISTARS under the European Union's Seventh Framework Programme (Grant Agreement no. 291294).
The NIKA dilution cryostat has been designed and built at the Institut N\'eel. In particular, we acknowledge the crucial contribution of the Cryogenics Group, and in particular Gregory Garde, Henri Rodenas, Jean Paul Leggeri. R. A. would like to thank the ENIGMASS French LabEx for funding this work. B. C. acknowledges support from the CNES post-doctoral fellowship program. E. P. acknowledges the support of grant ANR-11-BS56-015. 

\bibliography{cataSPIE}   
\bibliographystyle{spiebib}   

\end{document}